# High-repetition-rate solid tape target delivery system for ultra-intense laser-matter interaction at CLPU

**Michael Ehret** [1,*], **Diego de Luis** [1], **Jon Imanol Apiñaniz** [1], **Jose Luis Henares** [1], **Roberto Lera** [1], **José Antonio Pérez-Hernández** [1], **Pilar Puyuelo-Valdes** [1], **Luca Volpe** [2], **and Giancarlo Gatti** [1]

[1] *Centro de Láseres Pulsados (CLPU), Villamayor, Spain*
[2] *Universidad Politécnica de Madrid Escuela Técnica Superior de Ingeniería Civil, Madrid, Spain*

Correspondence*:
Michael Ehret, Centro de Láseres Pulsados, C/ Adaja 8, 37185 Villamayor, Spain
mehret@clpu.es

## ABSTRACT

The VEGA-3 laser system at the Centro de Láseres Pulsados (CLPU) delivers laser pulses up to $1\,\text{PW}$ at $1\,\text{Hz}$ repetition rate, focused to intensities up to $2.5 \times 10^{20}\,\text{W}\,\text{cm}^{-2}$. A versatile and compact targetry solution suitable for this repetition rate is presented. The system can operate in the challenging petawatt laser environment close to the laser-plasma interaction. Strips are spooled in a tape target system to deliver a solid density target to the laser focus for every shot. Results are presented for different tape materials and thicknesses. Experimental ion spectra are recorded by a Thomson Parabola Ion Spectrometer coupled with a scintillator screen; and an antenna array is used for the characterization of electromagnetic pulses. The results of both diagnostics show a good shot-to-shot stability of the system.

## 1 INTRODUCTION

Technological progress over the past decades [1, 2, 3, 4] led to the present availability of many PW-class Ti:Sa laser systems. When tightly focused on matter, the relativistic interaction regime opens up to laser-driven sources of charged-particle beams [5], neutrons [6] and X-rays [7]. Pulsed bright ion beams are generated from solid-density targets by well known mechanisms such as Target Normal Sheath Acceleration (TNSA), Radiation Pressure Acceleration (RPA) and others[8]. Many applications in Science and Technology benefit from laser-driven ion beams such as isotope production [9], positron emission tomography [10], ion beam microscopy [11], Particle-Induced X-ray Emission (PIXE) [12] as well as inertial confinement fusion [13].

At present, the most important technical difficulty of laser-driven sources is the repetition rate, as many difficulties arise when solid density targets are destroyed by laser and have to be replaced. A High-Repetition-Rate (HRR) operation $>1\,\text{Hz}$ is a major requirement for most of the above mentioned applications, which demands for HRR targetry solutions. Nowadays several approaches are pursued, e.g. generation of cryogenic ribbons [14], use of liquid jets [15, 16] or unwinding of a thin tape near the interaction position [17, 18, 19, 20, 21, 22, 23]. All of these schemes provide fast refreshing of target surface automatically and are able to deliver tens of thousands of targets in continuous operation under high-vacuum conditions. Tape targets stand out for their good vacuum compatibility.

This study presents the Tape Target System (TaTaS-PW) for $PW$-operation, developed at the Centro de Laseres PUlsados (CLPU). It is based on a previous iteration of the system operational up to $150\,TW$ [24]. The paper is organized as follows: first the experimental setup and methodology are presented. Secondly, experimental findings from HRR tape targets and individually aligned tape-like targets are compared. Thirdly, future steps towards elevated repetition rates are discussed. Finally, we conclude and comment on future experimental applications.

## 2  MATERIALS AND METHODS

Experiments for this work were conducted in the VEGA-3 laser facility at CLPU. There, the Ti:Sa laser pulse is amplified to an energy $E_L$ up to $30\,J$ per pulse, and compressed down to a duration $\tau_L$ of $28\,fs$. After compression, the short laser pulse is transported in high-vacuum of $1 \times 10^{-6}\,mbar$ to a $f = 2.5\,m$ off-axis parabola. The focal spot of $d_L = 12\,\mu m$ full-width at half-maximum (FWHM) is aligned onto the interaction plane, and maintained at constant size for this work. Laser energy and duration per pulse were not constant thorught this study, a controlled variation of these parameters was performed with a laser pulse duration between $28\,fs$ to $160\,fs$ and laser pulse energies on target ranging from $9\,J$ to $28\,J$. With this ranges the system delivers intensities from $1 \times 10^{19}\,W\,cm^{-2}$ to $2 \times 10^{20}\,W\,cm^{-2}$. The energy on target is extrapolated from calibrations recorded at low-energy, the focal spot at high energy is estimated to be the same as for low-energy measurements. For this experiment, $24\,\%$ of the energy on target are within in the first Airy disk at low power. The pulse duration is measured on-shot with a second-harmonic autocorrelator system that diagnoses the faint reflection from a thin pellicule positioned between parabola and focus.

Solid-density targets are placed in the laser focus position and tilted by $12.5°$ with respect to the laser axis to avoid reflection back towards laser amplifiers. As the VEGA laser pulse shows no pre-pulses capable of inducing a transparency or breakdown of the target [28], the main acceleration mechanism of charged particles is TNSA [29, 30] with the current laser and target parameters. In TNSA, a population of laser-heated relativistic electrons escapes the target and the successive potential dynamics leads to the formation of sheath fields which are capable of accelerating ionized surface contaminants up to several tens of $MeV\,u^{-1}$ [31]. A Thomson Parabola spectrometer (TP) with scintillator screen [32] is deployed to discern accelerated ions by energy and charge-to-mass ratio. The spectrometer aims at ions which are accelerated normal to the target surface; the scintillator screen is photographed for every shot and the image is post-processed with the online analysis tool IOSSR [33].

Broad-band Electro-Magnetic Pulses (EMP) are generated by laser-plasma interaction [25]. One source is a space charge field that builds up in the interaction chamber due to laser-driven target discharge [26], between the target at a positive potential and the vacuum-chamber walls at a negative potential. The magnitude and mode structure of electric and magnetic fields depend on the dimensions of the interaction chamber and the seed space charge distribution. Typical cavity resonances at VEGA-3 occupy the band of Very High Frequencies (VHF), notably hundreds of $MHz$ [27]. Commercially available calibrated magnetic- and electric-field antennas are used to capture electromagnetic waves inside the interaction chamber: the magnetic-field antenna MDF-9400 ($9\,kHz \pm 400\,MHz$, Aaronia AG) and the electric-field antenna OmniLOG-30800 ($300\,MHz \pm 8\,GHz$, AaroniaAG). The MDF-9400 is positioned $30\,cm$ behind the target following the laser axis, and the OmniLOG-30800 is $30\,cm$ behind the target and $30\,cm$ to the right (in laser direction), both oriented towards the interaction point. The signal is transported via calibrated double shielded SMA cables and through a floating feed-through. Waveforms are captured with an oscilloscope of $2\,GHz$ bandwidth and $10\,GS/s$ sampling rate. The grounding of shields in SMA cables is via the oscilloscope. All EMP signals were corrected for frequency dependent attenuation for all elements of the measurement circuit.

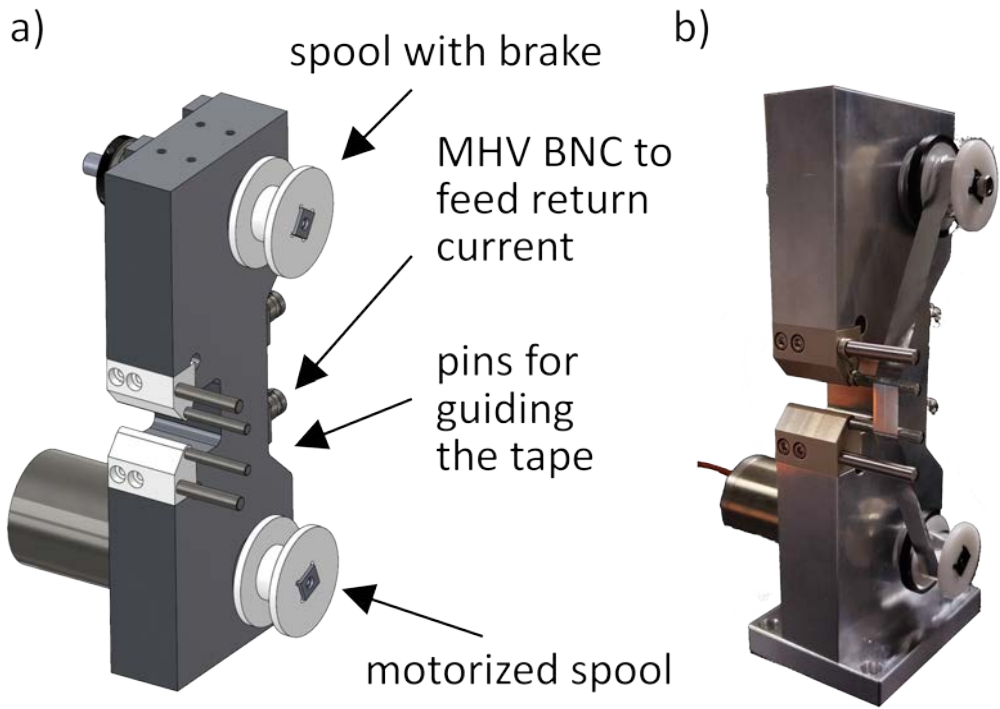

Figure 1. Tape target system for petawatt shots at VEGA-3: (a) technical details in the CAD illustration such as (i) the spool running over a break; (ii) the motorized spool; and (iii) the grounding pins keeping the tape in focal position; and (b) a photograph of the system. Note that pins are conductively connected to the ground via the MHV BNC connection in order to ensure grounding.

The tape target system TaTaS-PW is illustrated in Fig. 1. Spooling between laser shots, it transports several $\mu$m thick solid-density stripes of several mm width across the laser focal plane. The main frame of the TaTaS-PW system is made of a block of aluminum that incorporates (i) two shafts that house the tape spooling system, and (ii) two plastic teeth from PEEK that face each other across the laser-particle-plane and hold metallic pins which guide the tape over the laser focus plane.

(i) The spooling system comprises one spool propelled by a stepper motor Phytron VSS42 which pulls the tape from the other spool connected to a braking system of $2.5\,\mathrm{N\,cm}$ to keep the tape taut. Both spools are from dielectric material to isolate tape and main frame. In addition, the motor is isolated from the main frame to avoid destructive strong currents towards the motor controller unit, e.g. triggered by photo-ionization or stopped relativistic electrons on the main frame.

(ii) The pins at a horizontal distance of $\pm 5\,\mathrm{mm}$ ensure the tape stays in the laser-focus plane. The stability of the tape position normal to its surface is listed for different tape materials in Tab. 1. The root-mean-square (RMS) deviation of the surface positions are smaller than the Rayleigh length of VEGA-3, which is estimated to be $\approx 130\,\mu\mathrm{m}$ by beam diameter and focal length of the parabola. Choosing PEEK as material for the mounting structures of the pins ensures the isolation of the tape from the main frame. The pins can be grounded to any point via MHV-BNC connectors. For this work, grounding is ensured via the breadboard inside the vacuum chamber.

Table 1. RMS fluctuations of the tape target position normal to its surface, relative to the initially aligned focal spot position. Also given are the projections into the longitudinal direction relative to the laser (negative signs correspond to shifts towards the focusing parabola).

| Tape | Stable plane of Tape (normal) / [µm] | Longitudinal projection (laser) / [µm] |
| --- | --- | --- |
| Al | $-64 \pm 24$ | $-62 \pm 23$ |
| Kapton (89-K-l-ad) | $-3.0 \pm 3.3$ | $-2.9 \pm 3.2$ |
| Kapton-enforced Al (9-Al-e-K) | $-15 \pm 39$ | $-15 \pm 38$ |

The system allows versatile changes in operation: spool diameter and width can be adapted to required tapes; the controller unit allows free adjustment of parameters such as speed and acceleration; as well as spools and PEEK teeth can be exchanged swiftly in case of damage. The device can be used in diverse geometries needed for laser±matter interaction studies by providing a $310°$ free angle of view on the target in the equatorial plane, see Fig. 2.

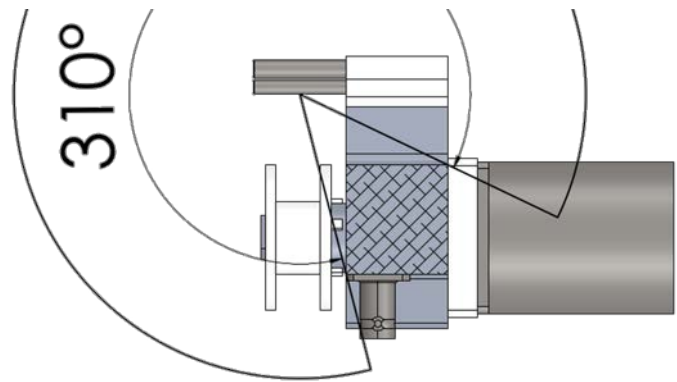

Figure 2. Top-view with a horizontal cut through the tape target system, with indication of the range that allows free field of view onto the interaction point on the tape.

## 3 RESULTS AND DISCUSSION

The tape system is capable of forwarding a $9\,\mu m$ thick Aluminium tape by $cm$ increments at $0.05\,Hz$ between laser shots at the same frequency every $20$ with $(0.70 \pm 0.12)\,PW$ in best compression. For this case, the proton cut-off energy in the TNSA spectrum is $(10.68 \pm 0.16)\,MeV$, see Fig. 3. The shot-to-shot standard deviation of the cut-off energy is $1.5\,\%$. The repetition rate was limited by a disconnection of the tape controller from the system for the duration of the shot and the successive re-connection, necessary to avoid perturbation by the EMP. Results are obtained in a burst with $39$ shots, the operation frequency extrapolates to $180$ shots per hour. Besides this demonstration of maximum performance, studies examined the performance with different types of tape and the effect of structured tapes.

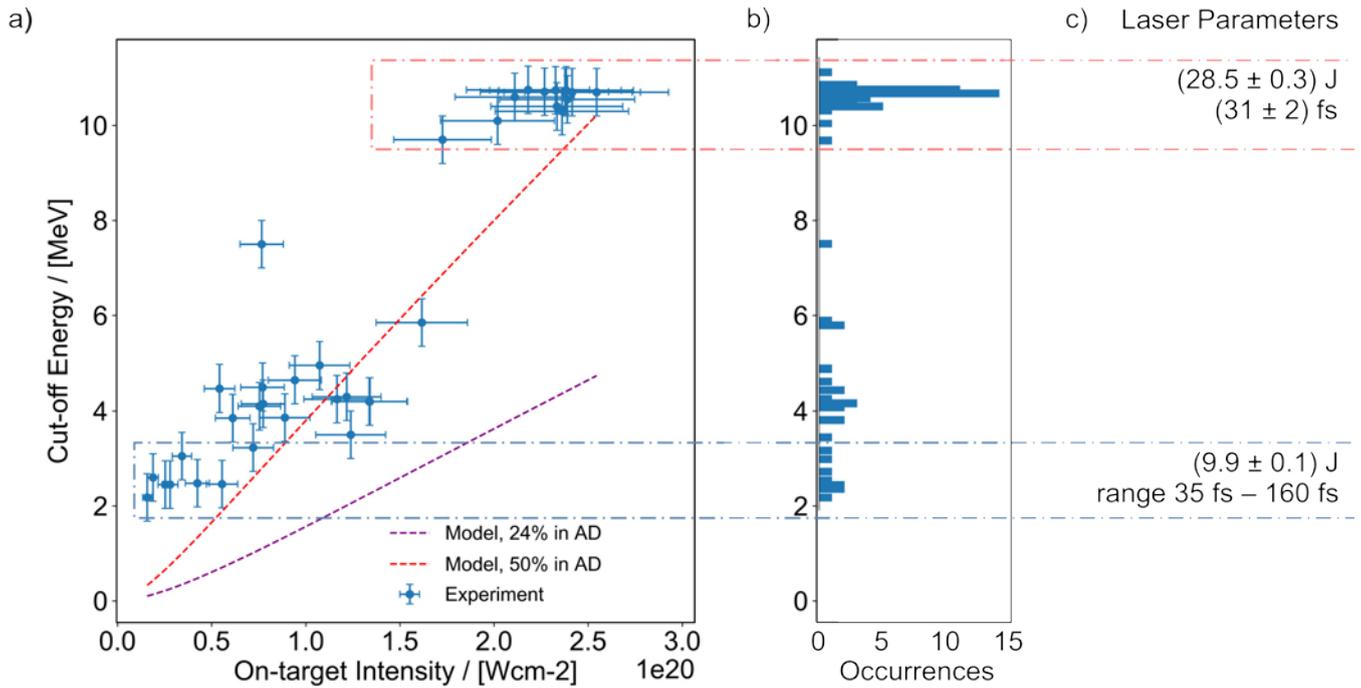

Figure 3. The spectral cut-off energy of accelerated proton beams is recorded for a fixed focal spot size but under variation of laser pulse energy and duration, results for Tape 9-Al-e-K. (a) The cut-off agrees with the modified Schreiber model [37]. (b) The histogram shows a narrow peak of $(10.68 \pm 0.16)$ MeV for a burst of $39$ shots at $0.05$ Hz with comparable laser parameters. (c) Even coarse variations of the laser pulse duration yield comparable cut-off energy when maintaining a constant pulse energy of $(9.9 \pm 0.1)$ J.

### 3.1 Study of Standard Tapes

Figure 3 shows all shots using the $10$ mm wide and $9$ µm thick Aluminium tape that is enforced with Kapton tape [1] on the edges (from now on named Tape 9-Al-e-K). Notably, a mm-thin stripe on each edge is used to smoothen the sliding of the tape over the inner pins and avoid rupture of the tape, see Fig. 4 (a). Results are compared to the modified Schreiber model [36, 37] in a wide range of laser pulse energies and durations. Model predictions roughly agree with measured values of the cut-off energy in this intensity regime of $1 \times 10^{19}$ W$^2$ cm$^{-1}$ to $1 \times 10^{20}$ W$^2$ cm$^{-1}$ if the fraction of laser energy within the first Airy disk (AD) is set to $50\%$. When applying the experimentally determined value of $24\%$, the model underestimates the cut-off energy by a factor of $2.5$. For model calculations, the electron beam divergence is estimated according to Debayle's model [38] and not measured, which may be the main cause of the discrepancies.

Tape 9-Al-e-K is compared to a $89$ µm thick Kapton tape with thin silicone adhesive on the rear side (from now on named Tape 89-K-l-ad). Tape 89-K-l-ad is shown in Fig. 4 (b), with circular ablation holes for low laser pulse energies and asymmetric shapes for higher energies. For comparable laser parameters, the evaporated volume of Tape 89-K-l-ad is larger than of Tape 9-Al-e-K. Both tapes are compared to a single-shot reference of $9$ µm thick Aluminium tape which is held by a target matrix. Different to shots on spooled tapes, the single-shot reference is aligned with respect to the laser focus for every shot. Figure 5 shows the proton cut-off energy for the respective shots. The three cases follow the same trend, between types cut-off energies do not vary in the margin of their uncertainty. The $89$ µm thick Kapton target allows for the same cutoff energy like a $9$ µm Aluminium target, for highest laser pulse energies in shots at best compression. The mechanisms involved should be part of future work, but the diagnosis of plasma parameters and ion acceleration is beyond the scope of this work. The comparability between Aluminium tape and single shot reference is an essential proof of the on-shot alignment stability of the system.

---

[1] UHV KAPTON TAPE SILICONE ADHESIVE, seller LewVac, dimensions 12.7mm width, 0.0889mm thickness and 32m length, purchased July of 2020

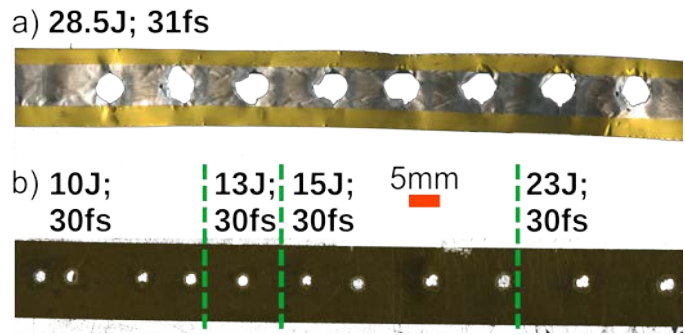

Figure 4. Tape types used during the experiment, respectively scanned after laser shots indicated with their laser pulse parameters. (a) Tape 9-Al-e-K with a mm-thin stripe sticked on each edge, and (b) Tape 89-K-l-ad pure Kapton tape with Silicon Adhesive on the rear side. (a) and (b) have the same scale.

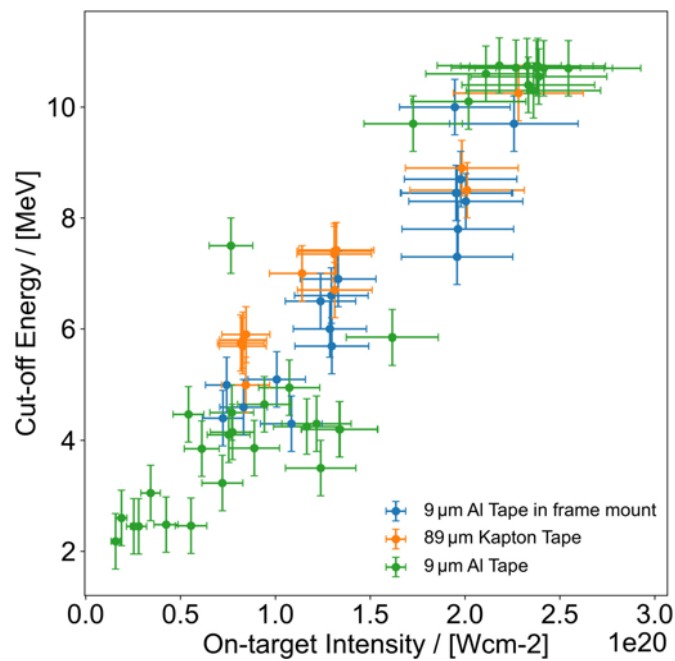

Figure 5. The spectral cut-off energy of accelerated proton beams is recorded for a fixed focal spot size but under variation of laser pulse energy and duration for three types of tape.

The stability of the cut-off energy of laser accelerated protons is a benchmark for the stability of the proton source. This metric is unavailable in experiments where ion beams are applied to secondary samples. The beam properties are altered by the interaction with samples, e.g. it is not possible to determine the amount of protons which loose all their energy inside the sample and are not transmitted. Here we investigate whether or not the EMP can be used for metrology to monitor the stability in such cases. The magnitude spectrum in the VHF band is sensitive to modifications in the experimental setup [27]. The evolution of the magnetic field amplitude for $39$ laser shots at $(28.5 \pm 0.3)\,\text{J}$ pulse energy and $(31 \pm 2)\,\text{fs}$ pulse duration on Tape 9-Al-e-K is shown in Fig. 6. The maximum and minimum amplitude of the time-domain measurement are stable, in accordance to the reproducible TNSA spectrum with a cut-off energy of $(10.68 \pm 0.16)\,\text{MeV}$ for the same shots (see Fig. 3). The EMP spectrum over the VHF band shows clear peaks in the magnitude spectrum. Their amplitude range is small around the mean value, which points to a reproducible mode structure.

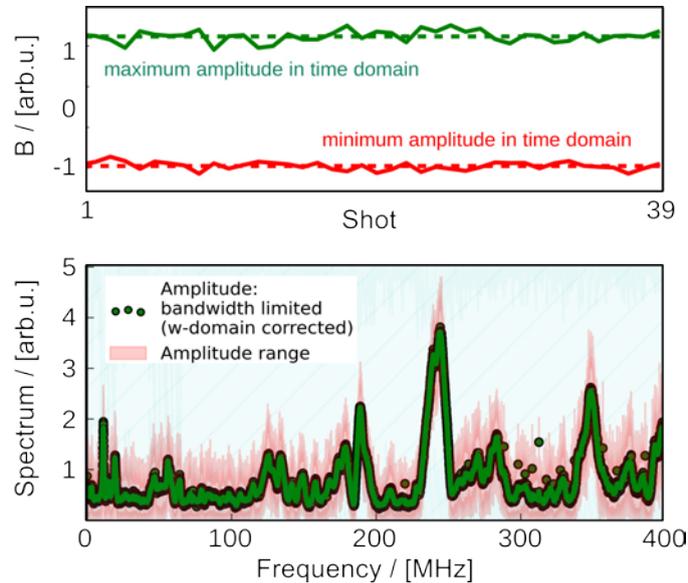

Figure 6. The maximum and minimum magnetic-field amplitude of the time-domain signal (top) and the time integrated average magnitude spectrum (bottom) of $39$ shots with the same laser parameters of $(28.5 \pm 0.3)\,\text{J}$ pulse energy and $(31 \pm 2)\,\text{fs}$ pulse duration on Tape 9-Al-e-K.

Figure 7 compares the signals of shots with energies ranging from $20\,\text{J}$ to $28\,\text{J}$ and and $(31 \pm 2)\,\text{fs}$ pulse duration. Shots marked with (a) are showing much lower non-zero amplitudes than shots marked with (b). Note that we did not detect any proton signal in the Thomson Parabola for shots (a) but for (b). Electric and magnetic field amplitudes are sensitive to miss-alignments. Which parts of the beamline were miss-aligned could not be determined for (a). Future characterisations should aim at a statistical analysis and mapping of EMP amplitudes depending on varying laser parameters and different layouts of the experimental setup.

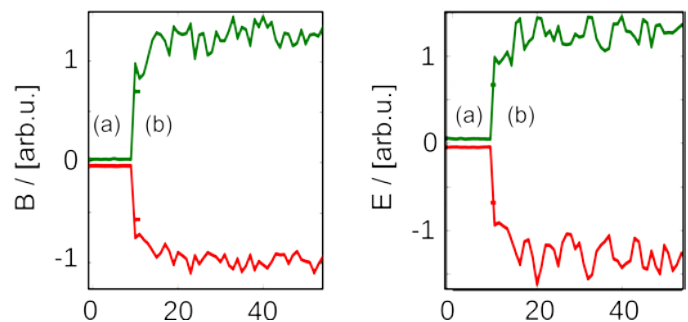

Figure 7. The maximum and minimum amplitude of the time-domain signal (left) for magnetic-fields and (right) for electric-fields of one run with shots on Tape 9-Al-e-K. Pulse energy and pulse duration vary from $20\,28$ and $(31 \pm 2)\,\text{fs}$. Shots (a) are biased by miss-alignments and shots (b) show sensitivity to laser parameters.

The TaTaS-PW system provides a stable targetry solution for HRR proton acceleration. Nevertheless, solid targetry produces a large amount of debris due to ablation. Mitigation strategies are addressed in the following with a study of different tape materials and geometries.

## 3.2 Study of Structured Tapes

Ablation holes in $9\,\mu m$ thick Aluminium are shown in Fig. 8 for different laser pulse energies at best compression. One notes that the laser punches a hole in the target by means of evaporation. The structure of the tape is weakened and the edges of the ablation zone curl, further increasing the size of the aperture. Thus, for low laser energies, the hole appears elliptical, which may be a matter of projection in the imaging system. For higher laser energies, there are clear deviations from the perfect circular shape. In the following, hole size refers to the size of the ablated zone taking into account curling effects. Hole diameters refer to diameters of circles which interpolate the true ablation zone, equating the respective volumes.

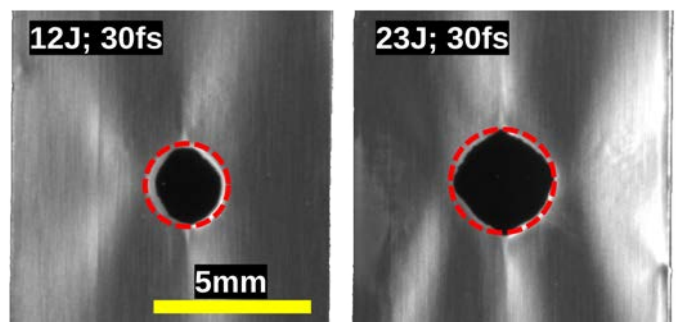

Figure 8. After the laser shot evaporated target material gives way to an ablation hole in the $9\,\mu m$ thick Aluminium tape, (left) for $12\,J$ and (right) for $23\,J$ laser pulse energy at best compression. Dashed lines indicate perfect circles (the circle on the right side is larger than the circle on the left side).

The ablation holes are compared to a model that predicts sizes of through-holes by assuming that the full heat transferred from the laser-heated electron population to the target drives the evaporation of target material [24]. All shots are conducted at best focus, and we compare experimental results and theoretical predictions in terms of laser pulse power versus hole diameters in Fig. 9. The model fails to predict hole diameters for cases where the laser pulse is stretched [39]. For shots at best compression the theoretical prediction is in the margin of uncertainty for $92\,\%$ of the data. With shots at a moderate peak-power of $\approx 400\,TW$ each shot evaporates $\approx 0.172\,mg$ of Aluminium. Continious HRR operation at $0.05\,Hz$ would cause debris of $1\,kg$ of Aluminium within $22.4\,d$.

Two strategies towards a reduction of debris and the control of the geometry of the ablated area are investigated, (i) with a multi-layer tape to dissipate the heat into a larger volume of low-Z material, and (ii) with micro-structured tapes to limit the region where electrons reflux through matter. Both targets are produced in tape shape and not tested in the tape target system, but pressed in a metal frame in order to allow precise individual alignment with respect to the laser focus.

The first approach is not successful. Fig. 10 shows in comparison to Fig. 8 that there is no difference to the hole diameter between single- and multi-layer tapes. The amount of ablated Aluminium is the same. The hole in the plastic layer appears to be un-altered. The layering may have been not well connected, a challenge that can be addressed with future experiments.

Structured targets proof to be effective for the reduction of debris, see Fig. 11. The hole diameter in structured tapes remains smaller than the diameter of holes in unstructured tapes. Note that structured tapes are cut in $6\,\mu m$ thick Aluminium, where holes are expected to be larger than for $9\,\mu m$ thick Aluminium as shown in [24]. The structuring reduces the amount of ablated material by $80\,\%$. Note that the closest edge of the micro-structure has a distance of $300\,\mu m$ to the laser interaction point to make sure the TNSA mechanism is not perturbed.

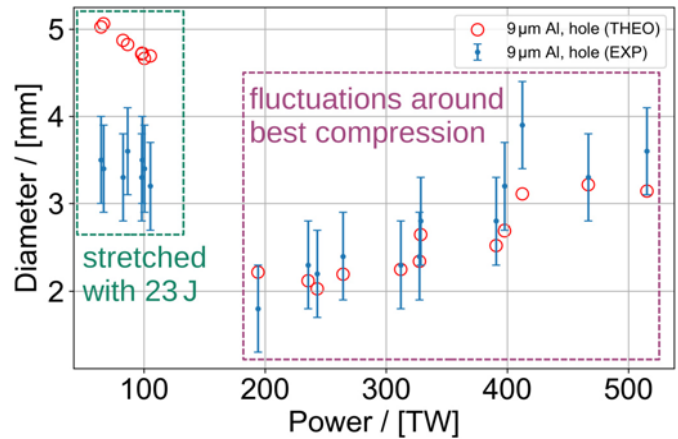

Figure 9. The diameter of the evaporation hole on a $9\,\mu m$ thick Aluminium tape under variation of the laser pulse power at best focus, compared are experimental results (EXP) and theoretical predictions (THEO). Highlighted are two groups of shots, (green) at $23\,J$ of laser power under variation of the laser pulse duration-, and (purple) around best compression under variation of the laser pulse energy.

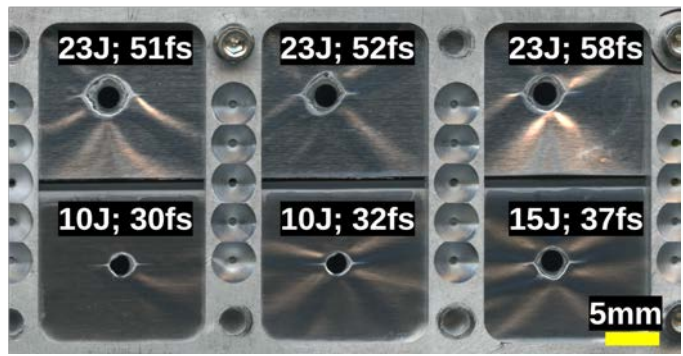

Figure 10. After the laser shot evaporated target material gives way to an ablation hole in the $9\,\mu m$ thick Aluminium tape, the $125\,\mu m$ thick Mylar layer that was pressed onto the Aluminium does not show clear signs of ablation.

Structuring allows for a precise control of the region in which material ablates. The circular hole in structured tapes remains intact which might prevent rupture of tapes when using thinner stripes of tape of reduced thickness. Such tapes are in demand because wide tapes can easily obscure direct lines of sight for side-view diagnostics, e.g. probe beams for plasma and pre-plasma dynamics, particle beam emission or Thomson scattering. Thinner foils are needed to not only access TNSA but also the RPA mechanism for ion acceleration.

## 4 CONCLUSIONS

We successfully implemented a tape target system that paves the way for high-repetition-rate experiments of relativistic laser-solid interaction. Stable ion acceleration via the TNSA mechanism is reported experimentally. The open geometry and modular configuration of the system allows for adaptation to specific experimental setups. A novel method for debris reduction via micro-structures is introduced.

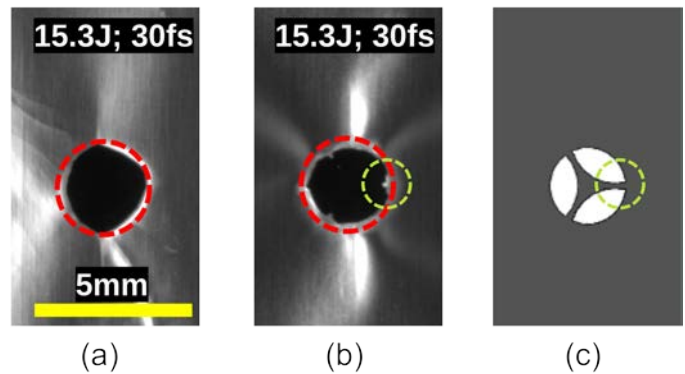

Figure 11. After the laser shot evaporated target material gives way to an ablation hole in (a) the $9\,\mu m$ thick Aluminium tape, and (b) the structured $6\,\mu m$ thick Aluminium tape. The essential difference is the amount of ablated material, (c) which is reduced in structured tapes by cut out (white) zones. Red circles have the same size. The design of the micro-structure is shown in (c) in comparable orientation to (b). Green circles highlight one arm of the micro structure, remnant after the laser shot in (b).

The stability of the system is evaluated in terms of ion spectra and EMP magnitude spectrum. The shot-to-shot standard deviation of the cut-off energy is $1.5\,\%$ for Kapton reinforced aluminium tapes of $9\,\mu m$ thickness. The EMP amplitudes in the VHF band in the time-domain provide a check for the stability of the interaction. The later method will gain importance when ion beams are applied to secondary sources and one can not check the proton spectrum for each shot.

The mechanical stability of the tape with respect to the initial alignment position is $(-15 \pm 38)\,\mu m$. Note that the Rayleigh length for the VEGA-3 focus ranges to approximately $130\,\mu m$.

Solid density targets are responsible for much debris. During the typical annual access time at CLPU of $18$ weeks, debris of $1\,kg$ could be issued from $9\,\mu m$ Aluminium foil when shot every $20\,s$ for $6\,h$ per day with $400\,TW$ at best focus. With respect to this large amount of debris, we demonstrate a reduction by $80\,\%$ by introducing micro-structures into the tape.

## CONFLICT OF INTEREST STATEMENT

The authors declare that the research was conducted in the absence of any commercial or financial relationships that could be construed as a potential conflict of interest.

## AUTHOR CONTRIBUTIONS

The author contributions are as follows: ME, DL, JA contributed equally to this work and share first authorship; ME wrote the first draft of the manuscript; DL, JA wrote sections of the manuscript; ME, DL, JH, RL contributed to conception and design of the study; ME, JAP performed the data curation and analysis; all authors were involved with underlying experimental work; all authors contributed to manuscript revision, read, and approved the submitted version.

## FUNDING

This work received funding from the European Union's Horizon 2020 research and innovation program through the European IMPULSE project under grant agreement No 871161 and from LASERLAB-EUROPE V under grant agreement No 871124. It benefited from funding from the Ministerio de Ciencia, Innovación y Universidades in Spain through ICTS Equipment grant No EQC2018-005230-P, further from grant PID2021-125389OA-I00 funded by MCIN / AEI / 10.13039/501100011033 / FEDER, UE and by ªERDF A way of making Europeº, by the ªEuropean Unionº and in addition from grants of the Junta de Castilla y León with No CLP263P20 and No CLP087U16.

## ACKNOWLEDGEMENTS

This work would not have been possible without the help of the laser- and the engineering team at CLPU. Special thanks to the workshop of CLPU.

## DATA AVAILABILITY STATEMENT

The raw data and numerical methods that support the findings of this study are available from the corresponding author upon reasonable request.